\documentclass[journal=jpcafh,manuscript=article]{achemso}

\usepackage{amsmath}
\usepackage{graphicx}
\usepackage{bm}
\usepackage{lscape}

\SectionNumbersOn

\author{Ryan Pederson}
\email{vgnc@vt.edu}

\author{Aleksander L. Wysocki}
\affiliation{Department of Physics, Virginia Tech, Blacksburg, Virginia 24061, United States}

\author{Nicholas Mayhall}
\affiliation{Department of Chemistry, Virginia Tech, Blacksburg, Virginia 24061, United States}

\author{Kyungwha Park}
\affiliation{Department of Physics, Virginia Tech, Blacksburg, Virginia 24061, United States}
\email{kyungwha@vt.edu}

\title{Multireference \emph{Ab Initio} Studies of Magnetic Properties of Terbium-Based Single-Molecule Magnets}

\begin{document}





\begin{abstract}
We investigate how different chemical environment influences magnetic properties of terbium(III) (Tb)-based single-molecule magnets (SMMs), using first-principles relativistic multireference methods. Recent experiments showed that Tb-based SMMs can have exceptionally large magnetic anisotropy and that they can be used for experimental realization of quantum information applications, with a judicious choice of chemical environment. Here, we perform complete active space self-consistent field (CASSCF) calculations including relativistic spin-orbit interaction (SOI) for representative Tb-based SMMs such as TbPc$_2$ and TbPcNc in three charge states. We calculate low-energy electronic structure from which we compute the Tb crystal-field parameters and construct an effective pseudospin Hamiltonian. Our calculations show that ligand type and fine points of molecular geometry do not affect the zero-field splitting, while the latter varies weakly with oxidation number. On the other hand, higher-energy levels have a strong dependence on all these characteristics. For neutral TbPc$_2$ and TbPcNc molecules, the Tb magnetic moment and the ligand spin are parallel to each other and the coupling strength between them does {\it not} depend much on ligand type and details of atomic structure. However, ligand distortion and molecular symmetry play a crucial role in transverse crystal-field parameters which lead to tunnel splitting. The tunnel splitting induces quantum tunneling of magnetization by itself or by combining with other processes. Our results provide insight into mechanisms of magnetization relaxation in the representative Tb-based SMMs.
\end{abstract}

\section{\label{sec:intro}Introduction}

Single-molecule magnets\cite{Chudnovsky1998,Friedman2010,Gao2015} (SMMs) are magnetic molecules typically composed of one or several transition metal or lanthanide ions surrounded by ligands. The key feature of SMMs is existence of inherent magnetic anisotropy due to interplay between the ligand crystal field (CF) and spin-orbit interaction (SOI). This feature allows one to explore interesting phenomena and potential applications of SMMs for magnetic information storage,\cite{Saywell2010,Guo2018} spintronics\cite{Romeike2006,Misiorny2007,Lopez2009,Burzuri2012,Schwobel2012} and quantum information processing.\cite{Aguila2014,Atzori2018,Leuenberger2001,Thiele2014,Shiddiq2016,Pedersen2016,Godfrin2017} Recently, it was reported that monometallic lanthanide-based SMMs can have an effective energy barrier of over 1000~cm$^{-1}$ with magnetic hysteresis above liquid nitrogen temperature.\cite{Guo2018,Goodwin2017} For reviews of lanthanide-based SMMs, see Refs.\citenum{Sessoli2009,Baldovi2012,Woodruff2013,Liddle2015,Wang2016}. Furthermore, since the first proposal of the implementation of Grover's algorithm into the prototype SMM Mn$_{12}$,\cite{Leuenberger2001} quantum bits (qubits) and quantum gates based on SMMs\cite{Thiele2014,Shiddiq2016,Pedersen2016,Godfrin2017} have been experimentally realized. One promising SMM candidate is a double-decker TbPc$_2$ family (Pc = phthalocyanine),\cite{Ishikawa2003} where Rabi oscillations and Grover's algorithm were experimentally implemented by tuning the nuclear spin states of the Tb ion with AC electric fields within single-molecule transistor set-ups.\cite{Thiele2014,Godfrin2017}

\begin{figure*}
\centering
\includegraphics[width=1.0\linewidth]{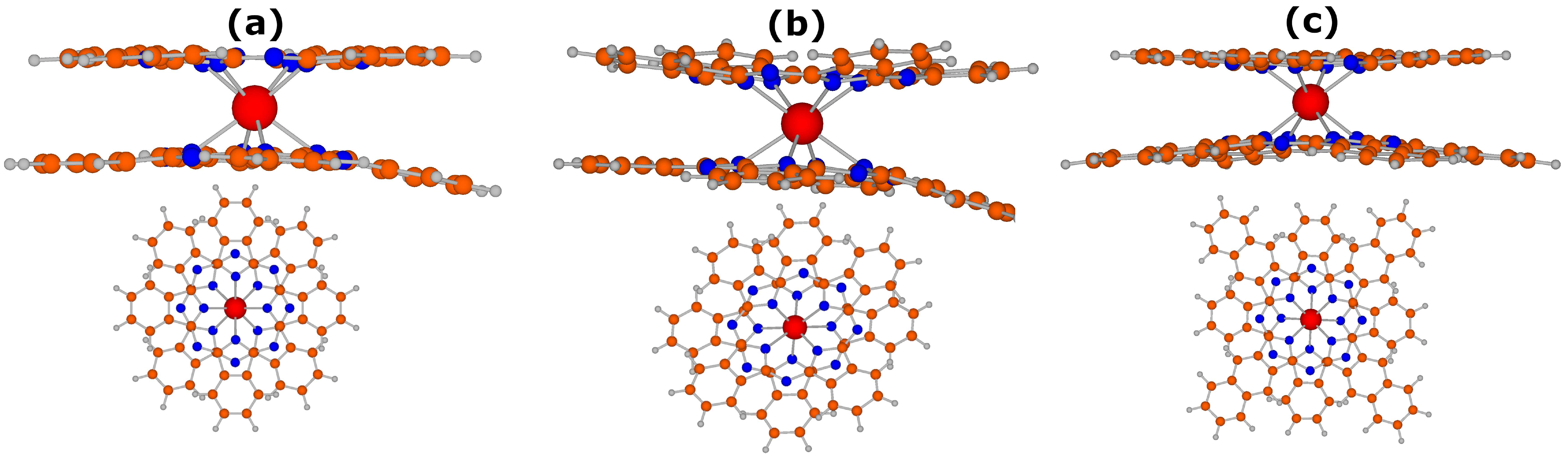}
\caption{Side and top views of experimental atomic structure of several TbPc$_2$-type molecules. (a) Neutral TbPc$_2$ with experimental geometry from Ref.~\citenum{Komijani2018} ({\bf M1}). (b) Anionic TbPc$_2$ with experimental geometry from Ref.~\citenum{Branzoli2009} ({\bf M5}). (c) Neutral TbPcNc with experimental geometry from Ref.~\citenum{Katoh2018} ({\bf M3}). Red, blue, orange, and gray spheres represent Tb, N, C, and H atoms, respectively.}
\label{Geometry}
\end{figure*}

SMM TbPc$_2$\cite{Ishikawa2003,Ishikawa2004} consists of a Tb$^{3+}$ ion sandwiched between two Pc ligands, as shown in Fig.~\ref{Geometry}a. Three different charge states were experimentally realized for TbPc$_2$ molecules such as anionic [TbPc$_2$]$^{-}$,\cite{Ishikawa2004b,Loosli2006,Takamatsu2007,Branzoli2009,Ganivet2013} (Fig. 1(b)) neutral [TbPc$_2$]$^{0}$,\cite{Katoh2009,Komijani2018} (Fig. 1(a)) and cationic [TbPc$_2$]$^{+}$.\cite{Takamatsu2007} In the latter case, however, X-ray crystallography data have not been reported yet. Recently, [TbPcNc]$^{0,+}$ molecules, where one of the Pc ligand rings was replaced by a larger Nc (naphthalocyaninato) ligand (Fig.~\ref{Geometry}c), were experimentally studied.\cite{Katoh2018} The synthesized TbPc$_2$ (TbPcNc) molecules have only approximate $D_{4d}$ ($C_{4v}$) symmetry. The degree of symmetry deviation varies with crystal packing, diamagnetic dilution molecules, or solvent molecules used in  synthesis processes.

Both TbPc$_2$ and TbPcNc molecules in different charge states were experimentally shown to exhibit SMM behavior.\cite{Ishikawa2003,Ishikawa2004b,Takamatsu2007,Katoh2009,Ganivet2013,Komijani2018,Loosli2006,Branzoli2009,Katoh2018} For the TbPcNc molecule, the measured effective energy barrier is in the range of 340-580 cm$^{-1}$ depending on oxidation number,\cite{Katoh2018} while for the TbPc$_2$ molecule, the measured barrier is in the range of 230-640~cm$^{-1}$ depending on oxidation number.\cite{Ishikawa2003,Ishikawa2004b,Takamatsu2007,Ganivet2013,Branzoli2009} There are no theoretical studies of the origin of this wide range of the energy barrier for TbPcNc or TbPc$_2$ molecules. Observed magnetization relaxation in SMMs may arise from combined contributions of different relaxation mechanisms such as quantum tunneling of magnetization (QTM), Raman process, Orbach process, hyperfine coupling, and/or intermolecular interaction.\cite{Aravena2018} In extracting the experimental barrier, it is often assumed that there are no intermolecular interactions and hyperfine interactions. The experimental barrier may also depend on magnetization relaxation mechanisms considered in the fitting of experimental data. Considering the complexity of the relaxation mechanisms and the assumption and ambiguity in the fitting process, it may be difficult to unambiguously determine the magnetization relaxation mechanisms solely from measurements. Furthermore, depending on experimental set-ups, additional environmental factors qualitatively affect the magnetic properties of the SMMs.

\begin{figure}
\centering
\includegraphics[width=1.0\linewidth]{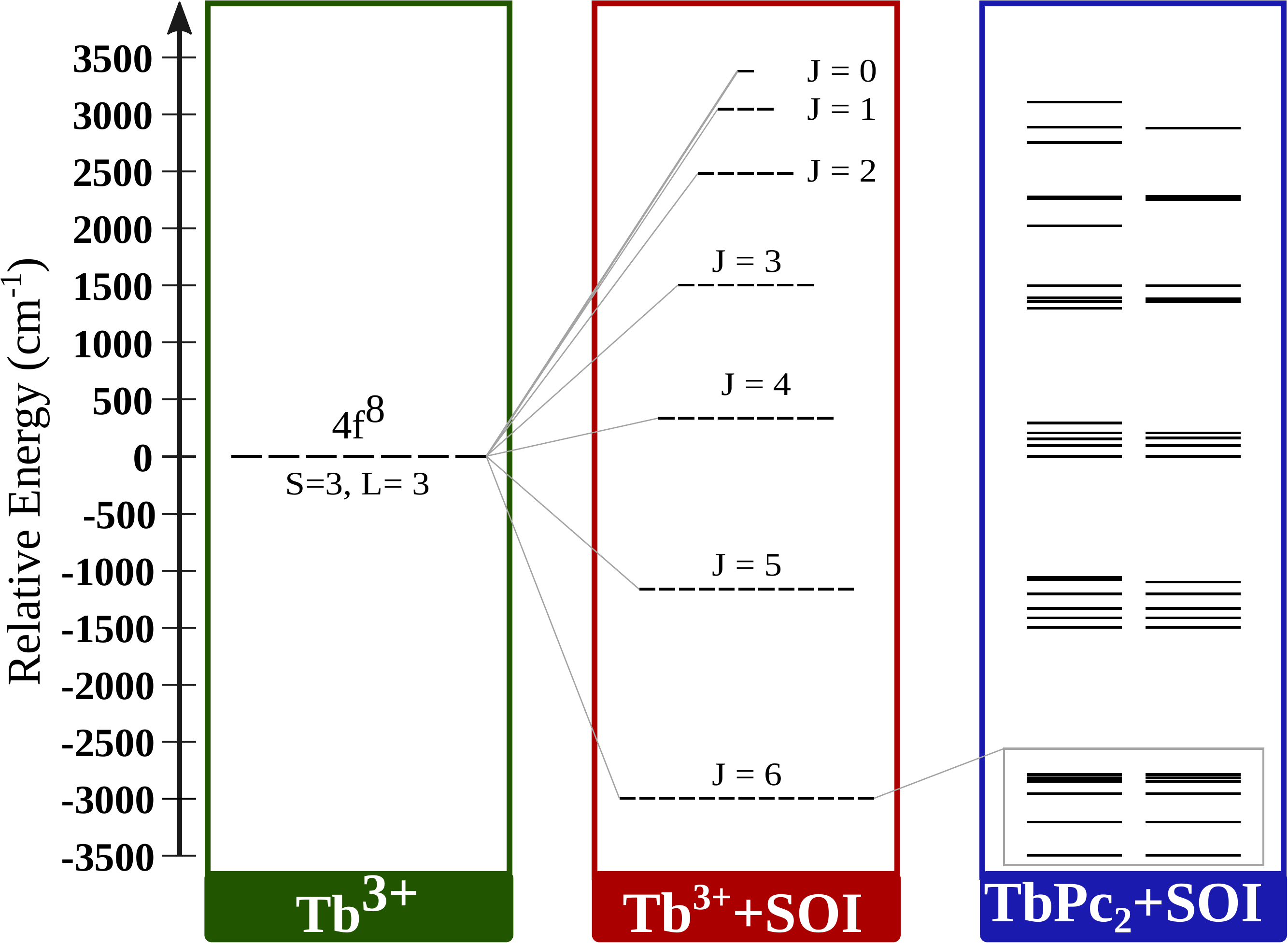}
\caption{Scalar relativistic $4f$ ($S=3$, $L=3$) energy level for an isolated Tb$^{3+}$ ion (left) splitted by SOI into $J$-multiplet structure (center). For TbPc$_2$ molecules the $J$-multiplets are further split by the CF of the Pc ligands (right). The energy levels in the boxed region correspond to the $J=6$ multiplet and are shown in Fig.~\ref{Spectrum} for charged TbPc$_2$-type molecules. The energy levels in the left and center panels are calculated for an isolated Tb$^{3+}$ ion while those on the right panel are obtained for {\bf M5} molecule (the relative energy scales are aligned such that the {\bf M5} lowest energy level is 3500 cm$^{-1}$ below the $S=3$, $L=3$ state).}
\label{TbLevels}
\end{figure}

In TbPc$_2$,\cite{Ishikawa2003,Ishikawa2004} the Tb$^{3+}$ ion has the 4f$^8$ electronic configuration. According to Hund's rules, its ground multiplet corresponds to spin, orbital and total angular momentum quantum numbers of $S=3$, $L=3$ and $J=6$, respectively. The unquenched orbital angular momentum gives rise to strong SOI. CF of Pc ligands in conjunction with the SOI splits the Tb $J$-multiplets (Fig.~\ref{TbLevels}), leading to large magnetic anisotropy. CF parameters are often referred to as magnetic anisotropy parameters. As shown in Fig.~\ref{TbLevels}, for low-energy $J$-multiplets, different $J$-multiplets are well separated from one another. In this case, assuming uniaxial magnetic anisotropy, total angular momentum projected onto the magnetic easy axis ($M_J$) remains a good quantum number. For a given $J$-multiplet, states with the same magnitude of $M_J$ are degenerate, whereas states with different $|M_J|$ values are split. Here we define magnetic anisotropy barrier (MAB) as the energy difference between the lowest and highest magnetic levels within the ground $J$-multiplet. Note that this barrier can differ from experimental effective energy barrier. The energy difference between the ground-state and the first-excited doublets for a given $J$-multiplet is referred to as zero-field splitting (ZFS).

For cationic or anionic TbPc$_2$ or TbPcNc SMMs, the Tb $J=6$ multiplet structure constitutes the entire low-energy spectrum. In this case, the transverse CF may split the two-fold degeneracy of nonzero $|M_J|$ levels by mixing states with different $M_J$. This phenomenon is referred to as tunnel splitting. For neutral TbPc$_2$ SMMs, one unpaired electron (with the spin $s=1/2$) is delocalized (or shared) within the two Pc ligands. This ligand spin interacts with the Tb magnetic moment by exchange coupling ($J_{\text{ex}}$) and doubles the number of low-energy levels. In addition, this extra electron makes neutral TbPc$_2$ a Kramers system, and irrespective of symmetry of the ligand field, it ensures at least two-fold degeneracy of all electronic levels enforced by time-reversal symmetry. Low-energy levels within the ground multiplet, ZFS, tunnel splitting, $J_{\text{ex}}$, and separation between the ground and first-excited multiplets are important energy scales that control the magnetic properties of TbPc$_2$ and TbPcNc SMMs. They play an important role in elucidation of magnetization relaxation mechanisms.

Despite the great interests and the experimental efforts and ambiguity, there are quite few computational studies of TbPc$_2$ molecule or its derivatives. In Ref.~\citenum{TbPc2-DFT} a neutral TbPc$_2$ molecule was investigated using density-functional theory (DFT) calculations with and without an on-site Coulomb repulsion $U$ term in the absence of SOI. However, SOI on Tb ion is much stronger than the ligand CF (Fig.~\ref{TbLevels}) and, therefore, it is imperative to include SOI for an even qualitative description of TbPc$_2$. Equally importantly, lanthanides atoms are known to have nearly degenerate electronic configurations demanding multireference treatments. The addition of the $U$ term alone does not suffice to describe the electronic structure and magnetic properties of the TbPc$_2$ molecule even qualitatively. Some multireference calculations of the TbPc$_2$ molecule using complete active space self-consistent field (CASSCF) method including SOI within restricted active space state-interaction (RASSI) have also been reported.\cite{marocchi2016relay,Ungur2017} In particular, in Ref.~\citenum{marocchi2016relay} a neutral TbPc$_2$ molecule was studied using a DFT-optimized structure when the molecule is adsorbed on a Ni substrate. In Ref.~\citenum{Ungur2017}, ZFS and MAB of anionic TbPc$_2$ were calculated with a particular experimental geometry.\cite{Branzoli2009} There are no multireference calculations of the magnetic properties of the TbPcNc molecule. Therefore, there is still a lack of multireference \emph{ab initio} studies of magnetic energy scales of TbPc$_2$-type SMMs as a function of charge state, type of ligand, and details of molecular geometry.

Here we investigate the effects of ligand and oxidation on the electronic and magnetic properties of the TbPc$_2$ and TbPcNc molecules in a gas phase, by using the CASSCF multireference method including SOI within RASSI. Using experimental geometries, we study the electronic levels characteristics and analyze the dependence of important magnetic energy scales on oxidation, ligand type and details of molecular structure. Furthermore, we construct an effective pseudospin Hamiltonian that includes Tb CF parameters and the Zeeman interaction (as well as exchange coupling between the Tb magnetic moment and the ligand spin for neutral molecules). The paper is structured as follows. Geometries of interest and methods used in this study are described in Sec.~\ref{sec:geometries} and Sec.~\ref{sec:methods}, respectively. Results of the TbPc$_2$ molecule in charged forms are followed by those for neutral forms in Sec.~\ref{sec:results}. We make conclusions in Sec.~\ref{sec:conc}.

\section{\label{sec:geometries}Geometries of Study}

We perform calculations for TbPc$_2$-type molecules with different charge states and ligand types for which experimental atomic structures are available. The use of experimental molecular geometries is preferred over theoretically optimized geometries since the latter, due to prohibitive computational costs, cannot include counter ions and solvent or dilution molecules which can be quite sizable. Indeed, the type of solvent molecules or diamagnetic dilution molecules as well as crystal packing, can significantly affect the ligand geometry of TbPc$_2$-type molecules. For example, two [TbPc$_2$]$^{-}$ molecules in Refs.~\citenum{Loosli2006} and \citenum{Branzoli2009} (referred to as {\bf M4} and {\bf M5} later) have significantly different geometries despite the same oxidation and ligand type (Table 1).

We consider the following six molecules (Table~\ref{StructuralParameters}): ({\bf M1}) neutral TbPc$_2$ with experimental geometry from Ref.~\citenum{Komijani2018}, ({\bf M2}) neutral TbPc$_2$ with experimental geometry from Ref.~\citenum{Katoh2009}, ({\bf M3}) neutral TbPcNc with experimental geometry from Ref.~\citenum{Katoh2018}, ({\bf M4}) anionic TbPc$_2$ with experimental geometry from Ref.~\citenum{Loosli2006}, ({\bf M5}) anionic TbPc$_2$ with experimental geometry from Ref.~\citenum{Branzoli2009}, ({\bf M6}) cationic TbPcNc with experimental geometry from Ref.~\citenum{Katoh2018}. As is a common practice, for all experimental geometries we correct the carbon-hydrogen bond length to 1.09~\AA~since this distance cannot be reliably extracted from X-ray measurements. Solvent molecules, diamagnetic dilution molecules, or counter ions are not included in our calculations.

\begin{figure}
\centering
\includegraphics[width=0.65\linewidth]{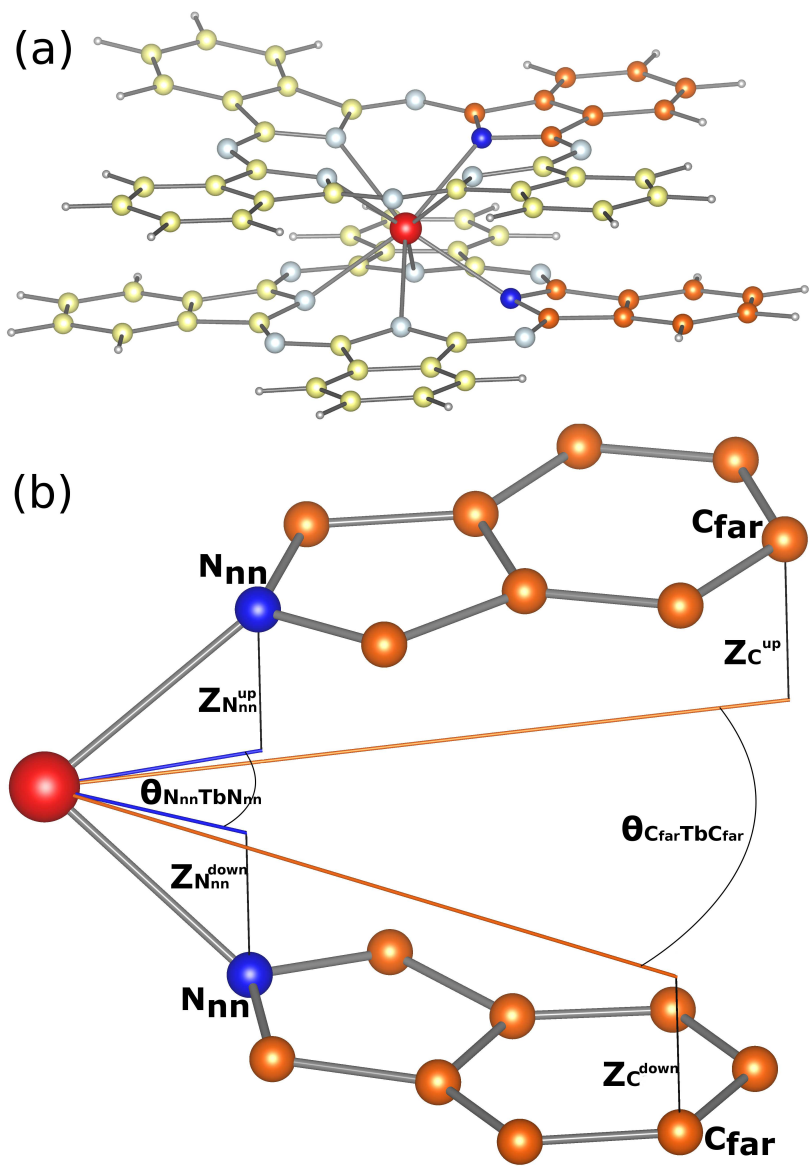}
\caption{Schematic molecular structure illustrating definitions of the structural parameters in Table \ref{StructuralParameters}. The highlighted region in (a) is zoomed in (b). The Tb ion is at the origin.}
\label{fig:StructPara}
\end{figure}

Figure \ref{Geometry} shows the atomic geometry of {\bf M1}, {\bf M5} and {\bf M3} molecules. The structure can be viewed in terms of the Tb ion sandwiched between two approximately flat Pc or Nc ligand planes. The ligands are rotated with respect to each other by roughly 45$^\circ$ angle. Each ligand has four roughly identical branches that form $\approx$90$^\circ$ angle with each other. Each branch starts with a nitrogen atom that is the nearest neighbor of the Tb ion and is denoted as N$_\text{nn}$ (Fig.~\ref{fig:StructPara}). Away from the Tb ion, N$_\text{nn}$ is connected to two intermediate carbon atoms which, in turn, are connected to a benzene-like carbon ring. Two carbon atoms from the ring which are the farthest from the Tb ion are denoted as C$_\text{far}$ (Fig.~\ref{fig:StructPara}). For Nc, C$_\text{far}$ atoms are further connected to an additional carbon ring. The neighboring branches are linked by bridging nitrogen atoms which form bonds with the intermediate carbon atoms.

\begin{landscape}
\begin{table*}
\centering
\caption{\textbf{Structural Parameters\textsuperscript{\emph{a}} for the Considered Experimental Geometries. The Errors Correspond to the Standard Deviation from the Average. Numbers in Parentheses Denote the Range of Observed Values where the Tb ion is at the origin. The Distances Are in \AA~and Angles in Degree}}
\begin{tabular}{lcccccc}
\hline
& [TbPc$_2$]$^0$ {\bf M1} & [TbPc$_2$]$^0$ {\bf M2}  & [TbPcNc]$^0$ {\bf M3}  & [TbPc$_2$]$^{-}$ {\bf M4} & [TbPc$_2$]$^{-}$ {\bf M5}
& [TbPcNc]$^{+}$ {\bf M6}       \\
\hline
Counter ion\textsuperscript{\emph{b}} &  n/a         & n/a & n/a & [TBA]$^+$$=$[N(C$_4$H$_9$)$_4$]$^{+}$ & [TBA]$^+$                 & (PF$_6$)$^{-}$ \\
Solvent or      & CH$_2$Cl$_2$ & n/a & CHCl$_3$ &   CH$_3$OH $\cdot$ H$_2$O      & 3([TBA]$^+$Br$^-$)       & 2 CH$_2$Cl$_2$ \\
diultion\textsuperscript{\emph{b}}        &              &     &          &                                & $\cdot$ 3$H_2$O          &                \\
Crystal packing\textsuperscript{\emph{c}} & $P_{nma}$ (62)& $P$2$_1$2$_1$2$_1$ (19) & $P$42$_1$2 (90) & $P_{na}$2$_1$ (33) & $P$2$_1$$c$ (56) & $P$42$_1$2 (90) \\
Symmetry\textsuperscript{\emph{d}} & $C_s$ & $C_1$ & $C_4$ & $C_1$ & $C_1$ & $C_4$ \\ \hline
$D_{\text{TbN}_{\text{nn}}}$\textsuperscript{\emph{e}}                        & 2.42$\pm$0.01  & 2.41$\pm$0.01  & 2.42$\pm$0.02  & 2.43$\pm$0.01  & 2.44$\pm$0.10  & 2.41$\pm$0.01  \\
$D_{\text{TbC}_{\text{far}}}$\textsuperscript{\emph{f}}                       & 6.79$\pm$0.05  & 6.82$\pm$0.04  & 6.85$\pm$0.03  & 6.88$\pm$0.07  & 6.91$\pm$0.41  & 6.85$\pm$0.03  \\
\hline
$\theta_{\text{N}_\text{nn}\text{TbN}_\text{nn}}$\textsuperscript{\emph{g}}   & (44.4,45.6)    & (40.6,49.5)    & (43.7,46.4)    & (43.3,46.8)    & (34.2,49.8)    & (43.4,46.6)    \\
$\theta_{\text{C}_\text{far}\text{TbC}_\text{far}}$\textsuperscript{\emph{h}} & (44.6,45.4)    & (40.7,49.9)    & (43.5,46.5)    & (42.8,46.8)    & (34.5,51.8)    & (43.4,46.6)    \\
\hline
$Z_{\text{N}_\text{nn}^\text{up}}$\textsuperscript{\emph{j}}                 & 1.41$\pm$0.00 & 1.39$\pm$0.01 & 1.39$\pm$0.00 & 1.41$\pm$0.02 & 1.41$\pm$0.03 & 1.38$\pm$0.00 \\
$Z_{\text{N}_\text{nn}^\text{down}}$\textsuperscript{\emph{i}}                  & -1.41$\pm$0.01  & -1.40$\pm$0.02  & -1.41$\pm$0.00  & -1.41$\pm$0.01  & -1.42$\pm$0.04  & -1.40$\pm$0.00  \\
$Z_{\text{C}^\text{up}}$\textsuperscript{\emph{k}}                  & (1.49,1.66)    & (1.51,2.00)    & (1.51,1.78)    & (1.53,2.68)    & (1.46,2.45)    & (1.50,1.81)    \\
$Z_{\text{C}^\text{down}}$\textsuperscript{\emph{l}}                & (-2.38,-1.47)  & (-2.26,-1.49)  & (-2.12,-1.57)  & (-2.83,-1.52)  & (-2.99,-1.36)  & (-2.19,-1.55)  \\
$Z_{\text{C}^\text{ave}_\text{down}}$;$Z_{\text{C}^\text{ave}_\text{up}}$\textsuperscript{\emph{m}} &
-1.70;1.59 & -1.68;1.69 & -1.54;1.66 & -1.90;1.92 & -1.84;1.91 & -1.57;1.67 \\
Reference & ~\citenum{Komijani2018} & ~\citenum{Katoh2009} & ~\citenum{Katoh2018} & ~\citenum{Loosli2006} & ~\citenum{Branzoli2009} & ~\citenum{Katoh2018} \\
\hline
\end{tabular}
\\
\raggedright
\textsuperscript{\emph{a}} See Fig.~\ref{fig:StructPara};
\textsuperscript{\emph{b}} Counter cations or anions, solvent molecules, or diamagnetic dilution molecules that separate individual SMMs from each other;
\textsuperscript{\emph{c}} The type of crystal packing with space group number in parenthesis;
\textsuperscript{\emph{d}} Point group of each individual SMM;
\textsuperscript{\emph{e}} Average bond length between Tb and N$_\text{nn}$ atoms;
\textsuperscript{\emph{f}} Average bond length between Tb and C$_\text{far}$ atoms;
\textsuperscript{\emph{g}} Average N$_\text{nn}$-Tb-N$_\text{nn}$ angle (with all three atoms projected on the same ligand plane defined by four N$_\text{nn}$ atoms) where the nitrogen atoms are the closest N$_\text{nn}$ from different ligands;
\textsuperscript{\emph{h}} Average C$_\text{far}$-Tb-C$_\text{far}$ angle (with all three atoms projected on the same ligand plane defined by four N$_\text{nn}$ atoms) where carbons are right and left C$_\text{far}$ atoms that belong to the closest branches from different ligands;
\textsuperscript{\emph{i}} Average vertical coordinate of N$_\text{nn}$ atoms from upper ligand plane;
\textsuperscript{\emph{j}} Average vertical coordinate of N$_\text{nn}$ atoms from lower ligand plane;
\textsuperscript{\emph{k}} Minimum and maximum vertical coordinates among
all C atoms from upper ligand plane;
\textsuperscript{\emph{l}} Minimum and maximum vertical coordinates among
all C atoms from lower ligand plane;
\textsuperscript{\emph{m}} Average vertical coordinate of all C atoms from
lower and upper ligand planes.
\label{StructuralParameters}
\end{table*}
\end{landscape}

In order to characterize the experimental geometries we introduce several structural parameters (see Table \ref{StructuralParameters} and Fig.~\ref{fig:StructPara}). The analysis of these parameters allows us to quantify deviations of the molecular structure from the ideal $D_{4d}$ (or  $C_{4v}$ for TbPcNc) symmetry induced by counter cations/anions, solvent molecules, diamagnetic dilution molecules, or crystal packing. For {\bf M1}, {\bf M3}, and {\bf M6} molecules, both $\theta_{\text{N}_\text{nn}\text{TbN}_\text{nn}}$ and $\theta_{\text{C}_\text{far}\text{TbC}_\text{far}}$ angles are close to 45$^\circ$ which indicates that the four-fold symmetry is approximately preserved. To a somewhat lesser degree, this is also true for {\bf M2} and {\bf M4} molecules. This is consistent with a small standard deviation of the $D_{\text{TbN}_{\text{nn}}}$ and $D_{\text{TbC}_{\text{far}}}$ bond lengths for these molecules. On the other hand, for {\bf M5} molecule, we have strong deviations from the four-fold symmetry as reflected in the fact that both the angles and the $\theta_{\text{N}_\text{nn}\text{TbN}_\text{nn}}$ and $\theta_{\text{C}_\text{far}\text{TbC}_\text{far}}$ as well as the $D_{\text{TbN}_{\text{nn}}}$ and $D_{\text{TbC}_{\text{far}}}$ bond lengths vary significantly. For all considered geometries, we observe curving of the carbon parts of the ligand planes away from each other (this effect can be also seen in Fig. \ref{Geometry}). The curving can be different for different ligand branches and it is most pronounced for the {\bf M5} molecule. The strong deviations of the {\bf M5} geometry from the ideal TbPc$_2$ structure is likely a result of bulky diamagnetic dilution molecules used in the synthesis process.\cite{Branzoli2009}

\section{\label{sec:methods}Methods}

The multireference calculations are performed using the Molcas quantum chemistry code (version 8.2).\cite{Molcas} Scalar relativistic effects are included based on the Douglas-Kroll-Hess Hamiltonian\cite{Douglass1974,Hess1986} using relativistically contracted atomic natural orbital (ANO-RCC) basis sets.\cite{Widmark1990,Roos2004} In particular, polarized valence triple-$\zeta$ quality (ANO-RCC-VTZP) is used for the Tb ion, polarized valence double-$\zeta$ quality (ANO-RCC-VDZP) is used for the nitrogen and carbon atoms, and valence double-$\zeta$ quality (ANO-RCC-VDZ) is used for the hydrogen atoms. Such a choice of the basis set is made to maintain a high accuracy and to not exceed computational capabilities. More details on the basis set dependence are discussed in Tables S4 and S5 in Supporting Information.

\begin{figure}
\centering
\includegraphics[width=0.9\linewidth]{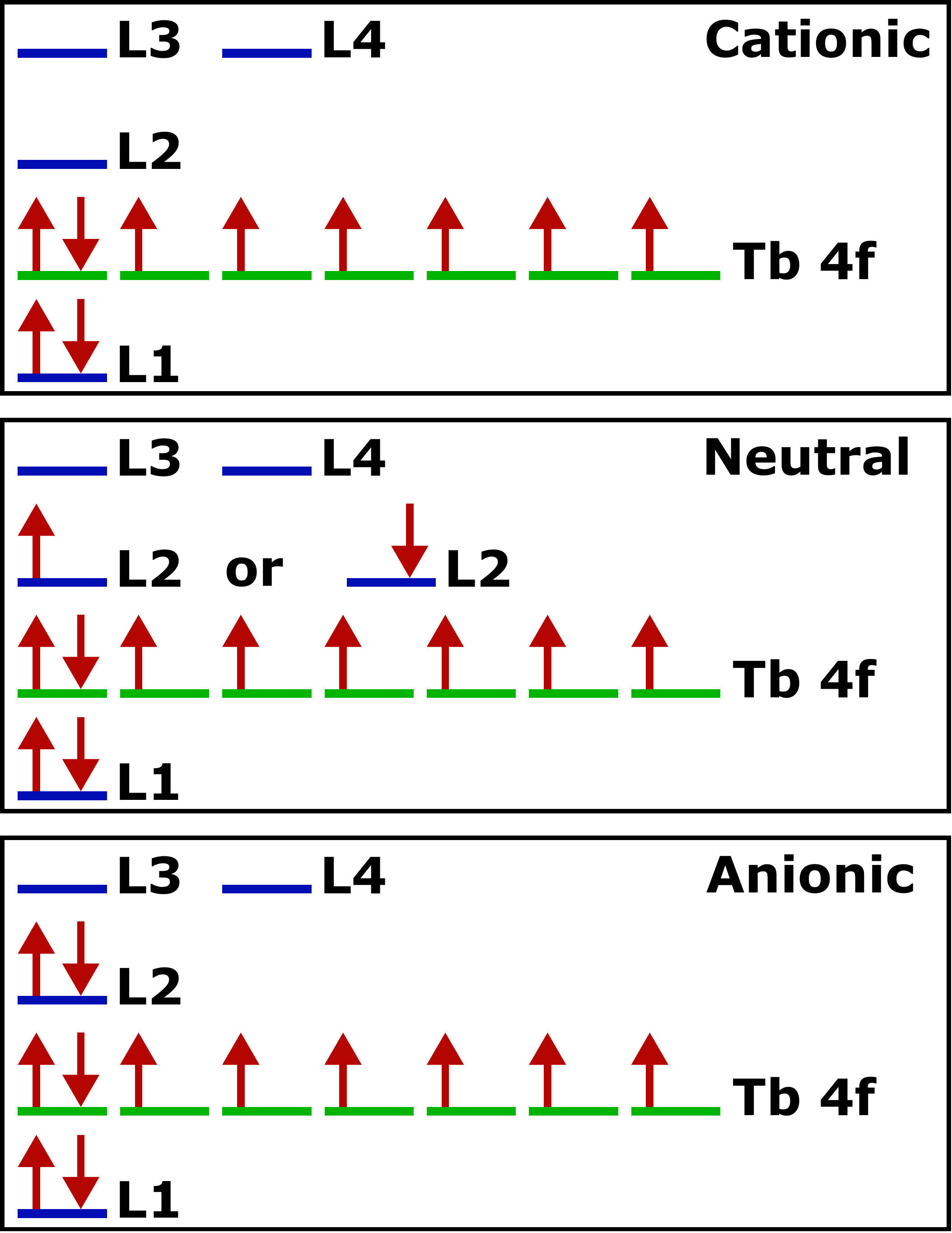}
\caption{Schematic illustration of the active space used in the CASSCF calculations for different charge states. L1, L2, L3, and L4 denote four ligand orbitals that lie close to the HOMO-LUMO gap. The figure shows nominal occupation of the orbitals.}
\label{ActiveSpace}
\end{figure}

First, in the absence of SOI, for a given spin multiplicity, the spin-free eigenstates are obtained using state-averaged CASSCF method.\cite{Roos1980,Siegbahn1981} The valence electronic configurations of Tb$^{+3}$ ion consists of eight electrons at 4$f$ orbitals which must be included in the active space. It would be desirable to include all ligand $\pi$/$\pi^\star$-type orbitals as well. This is, however, computationally prohibitive and, therefore, only several near-in-energy ligand orbitals are included in the active space. In order to identify such ligand orbitals, we consider molecules in the cationic state and perform CASSCF calculations with eight electrons and seven Tb 4$f$-type orbitals in the active space. We find that HOMO (L1) is always energetically separated ($\sim$0.1 a.u.) from other occupied ligand orbitals. For unoccupied orbitals we find that LUMO-1 (L2) as well as nearly degenerate LUMO-2 (L3) and LUMO-3 (L4) are separated from higher unoccupied states. Therefore, we include L1, L2, L3, and L4 ligand orbitals in the active space. Altogether, as illustrated in Fig.~\ref{ActiveSpace}, we consider eleven active orbitals with ten, eleven, and twelve active electrons for cationic, neutral, and anionic molecules, respectively. The effect of the choice of the active space on the final results is discussed in Supporting Information (Tables S7 and S8).

In the case of charged molecules we consider only the $S=3$ configuration since other spin configurations such as $S=2$, $S=1$, and $S=0$ are much higher in energy and their inclusion changes the energy levels of the ground $J$-multiplet only by about a few cm$^{-1}$. See Table S6 in Supporting Information for details. For neutral molecules, depending on whether the ligand spin ($s=1/2$) is parallel or antiparallel to the Tb spin $S=3$, we have two possible values of the total spin of the molecule: $S_{\text{tot}}=7/2$ or $S_{\text{tot}}=5/2$ (see Fig.~\ref{ActiveSpace}). We consider both $S_{\text{tot}}$ values since they lie close in energy. For a given spin configuration, we evaluate seven lowest spin-free states (roots) that correspond to different configurations of eight electrons in Tb 4$f$-type orbitals with Tb spin of $S=3$. These seven spin-free states are used in the state-averaged procedure.

In the next step, we include SOI, within the atomic mean-field approximation,\cite{Hess1996} in the space of aforementioned spin configurations and spin-free eigenstates, using the RASSI method.\cite{rassi} With SOI, all possible $J$-multiplets from the addition of ${\mathbf L}$ and ${\mathbf S}$ are generated as illustrated in Fig.~\ref{TbLevels}. For the calculation of the CF parameters, we use the methodology implemented in the SINGLE\_ANISO\cite{chibotaru2012ab} module of the Molcas code.

The technique from Ref.~\citenum{chibotaru2012ab} that we use for the charged molecules cannot be directly applied to the neutral molecules. Indeed, for lanthanides this method finds CF parameters and $g$-tensor elements for a given $J$-multiplet using lowest $(2J+1)$ \emph{ab initio} eigenvalues and the corresponding eigenfunctions. For neutral molecules, however, the low-energy levels do not only originate solely from the ground $J$-multiplet but also they involve the unpaired ligand electron spin-flip states. Both types of excitations are entangled and there is no obvious way how to extract multiplet levels and their wave functions. In fact, the entanglement is essential for formation of Kramers doublet. In order to circumvent this problem, we take the experimental geometries of the neutral molecules and consider them in the cationic state for the calculation of the Tb CF parameters. In this way we remove the unpaired ligand electron so that the calculated low-energy spectrum can be put into correspondence with the Tb $J=6$ multiplet and the CF parameters as well as the $g$-tensor elements can be calculated. Assuming that the unpaired ligand electron has a small contribution to the Tb CF and $g$-tensor, these parameters as well as the exchange coupling constant can be used in effective pseudospin
Hamiltonian for the neutral molecules.

\section{\label{sec:results}Results and Discussion}

In each subsection, we present the calculated magnetic energy levels obtained from CASSCF-RASSI-SOI method and construct the effective pseudospin Hamiltonian with the calculated CF parameters and $g$ tensor. We then compare with relevant theoretical results and experimental data when they are available.

\subsection{\label{sec:charged}Charged molecules}

\begin{figure}
\centering
\includegraphics[width=0.85\linewidth]{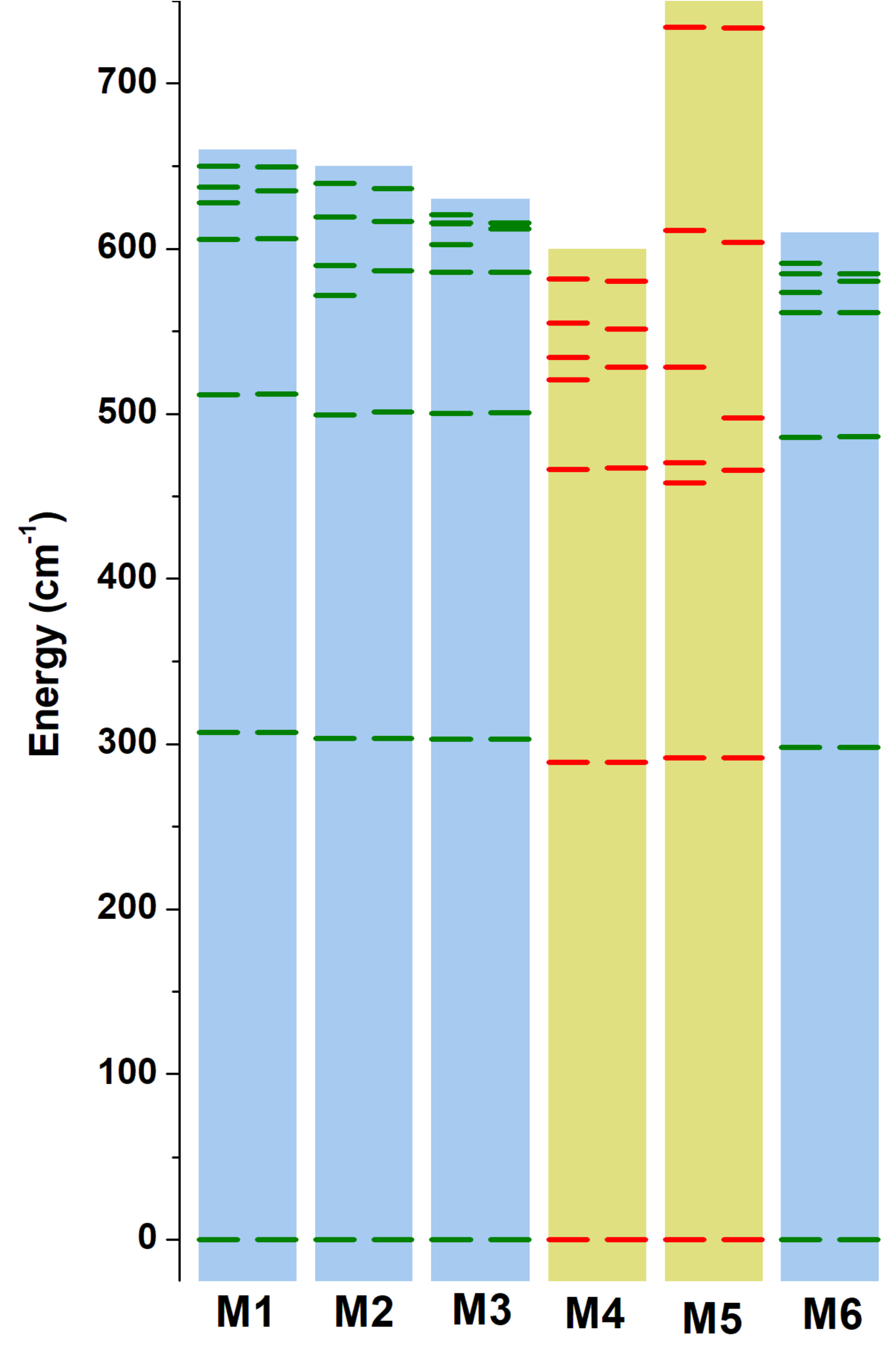}
\caption{Calculated low-energy ($J=6$) spectrum for different charged TbPc$_2$-type molecules. For {\bf M1}, {\bf M2}, and {\bf M3} molecules calculations are done in the cationic state with neutral-state geometries.}
\label{Spectrum}
\end{figure}

\begin{table}
\centering
\caption{\textbf{Calculated Magnetic Energy Scales for Different TbPc$_2$-Type SMMs in Units of cm$^{-1}$}}
\begin{tabular}{lcccccc}
\hline
                       & {\bf M1} & {\bf M2}   & {\bf M3}   & {\bf M4}    & {\bf M5}    & {\bf M6}    \\
\hline
$E_{\text{TS}}^{1,2}$\textsuperscript{\emph{a}}    & n/a  & n/a  & n/a  & 0.000 & 0.007 & 0.000 \\
$E_{\text{TS}}^{3,4}$\textsuperscript{\emph{b}}    & n/a  & n/a  & n/a  & 0.015 & 0.090 & 0.000 \\
$E_{\text{TS}}^{5,6}$\textsuperscript{\emph{c}}    & n/a  & n/a  & n/a  & 0.530 & 7.969 & 0.499 \\
$\Delta E_{\text{ex}}$\textsuperscript{\emph{d}}   & 8.2  & 6.6  & 5.1  & n/a   & n/a   & n/a   \\
$J_{\text{ex}}$\textsuperscript{\emph{e}}           & 0.8  & 0.6 & 0.6  & n/a    & n/a   & n/a   \\
$E_{\text{ZFS}}$\textsuperscript{\emph{f}}          & 308  & 305  & 308  & 289   & 292   & 298   \\
$E_{\text{MAB}}$\textsuperscript{\emph{g}}          & 658  & 639  & 622  & 582   & 734   & 592   \\
$\Delta E_J$\textsuperscript{\emph{h}}               & 2083 & 2068 & 2068 & 2036  & 2045  & 2049  \\
\hline
\end{tabular}
\\
\raggedright
\textsuperscript{\emph{a}} Tunnel splitting for the ground-state quasi-doublet;
\textsuperscript{\emph{b}} Tunnel splitting for the first excited quasi-doublet;
\textsuperscript{\emph{c}} Tunnel splitting for the second excited quasi-doublet;
\textsuperscript{\emph{d}} Energy difference between states with parallel and antiparallel orientation of the Tb angular momentum and the ligand spin;
\textsuperscript{\emph{e}} Exchange coupling between the Tb angular momentum and the ligand spin;
\textsuperscript{\emph{f}} ZFS;
\textsuperscript{\emph{g}} MAB;
\textsuperscript{\emph{h}} Separation between the ground and first-excited $J$ multiplets.
\label{Scales}
\end{table}

For charged TbPc$_2$-type molecules ({\bf M4}, {\bf M5} and {\bf M6}), we find that among eleven active molecular orbitals, seven are Tb 4$f$-type. These orbitals are similar for all considered molecules and are shown in Fig. S1 in Supporting Information. The occupation of each 4$f$-type orbital is approximately 1.14 which corresponds to about eight electrons occupying these orbitals (note that this is a result of state-averaged calculations). The remaining four active orbitals (L1-L4) are ligand orbitals and are shown in Figs. S5-S7 in Supporting Information. For cationic {\bf M6} molecule, L1 is almost doubly occupied while L2-L4 are almost empty. All ligand orbitals arise mostly from inner C atoms and they have in-plane symmetry due to $C_4$ symmetry of the molecule (Table 1). There is, however, asymmetry between orbitals in the top Pc and the bottom Nc ligands. For anionic {\bf M4} molecule, the occupation of nominally filled L1 and L2 orbitals is somewhat lower than two while the nominally empty L3 and L4 orbitals have a significant occupation (Fig. S5). The effect is even stronger for the {\bf M5} molecule (Fig. S6) and it suggests sizable correlations. The orbitals have significant in-plane asymmetry as well as asymmetry between the two Pc planes. This is a consequence of deviations from ideal $D_{4d}$ symmetry as illustrated by large spread and uneven distribution of $Z_{\text{C}^\text{up}}$ and $Z_{\text{C}^\text{down}}$ structural parameters (Table 1). 

For the charged molecules, the low-energy spectrum is largely determined by Tb atomic levels which are split in the ligand CF. As listed in Table \ref{Scales}, the separation between the ground and the first-excited $J$-multiplet, $\Delta E_J$, is almost a factor of three larger than the MAB of the ground $J$-multiplet, $E_{\text{MAB}}$. Thus, we can project the lowest $J$ multiplet onto a multiplet of effective pseudospin $\tilde{J}=6$. Figure~\ref{Spectrum} shows the thirteen lowest calculated energy levels corresponding to the $J=6$ ground multiplet for all considered charged molecules (numeric data together with spin-free energies are provided in Tables S1 and S2 in Supporting Information). In order to elucidate the properties of these levels we construct an effective pseudospin Hamiltonian

\begin{equation}
\hat{H}=\sum_{k=2, {\mathrm{even}}} \sum_{q=-k}^{k} B_k^q \hat{O}_k^q(\hat{\mathbf{J}}) -\mu_B\mathbf{B}\cdot \mathbf{g}\cdot\hat{\mathbf{J}},
\label{H1}
\end{equation}
where $\hat{O}_k^q(\hat{\mathbf{J}})$ are $k$-th rank extended Stevens operators\cite{Rudowicz1985} and $B_k^q$ are corresponding CF parameters ($q=-k,...,0,...,k$). Here $\hat{\mathbf{J}}$ is the Tb pseudospin operator, and $\hat{O}_k^{-q}(\hat{\mathbf{J}})=[\hat{O}_k^q(\hat{\mathbf{J}})]^{\star}$, where $\star$ means complex conjugate. For example, second-rank Stevens operators are $\hat{O}_2^0=(3J_z^2 - J(J+1)I)$, $\hat{O}_2^{1}=(J_z (J_{+} + J_{-}) + (J_{+} + J_{-}) J_z)/4$ and $\hat{O}_2^{2}=(J_+^2 + J_-^2)/2$, where $J_{\pm}$ are raising and lowering operators and $I$ is a 2$\times$2 identity matrix. Higher-rank Stevens operators are listed in Table S9 in Supporting Information. Here $\hat{O}_k^{q=0}$ terms represent uniaxial or diagonal contributions, while $\hat{O}_k^{q \neq 0}$ terms are transverse or off-diagonal contributions. Time-reversal symmetry enforces only even integer $k$ in the summation. Molecular symmetry dictates allowed nonzero $B_k^q$ values. When the second-order uniaxial term, $\hat{O}_2^0$, is dominant, ZFS and MAB are approximated to be $3|B_2^0|(2J-1)$ and $3|B_2^0|J^2$, respectively, for
integer $J$. The second term in Eq.~(\ref{H1}) is the Zeeman interaction with the $g$-tensor, $\mathbf{g}$.

\begin{table}
\centering
\caption{\textbf{Calculated Principal Values\textsuperscript{\emph{a}} of the $g$-Tensor for Different TbPc$_2$-Type SMMs}}
\begin{tabular}{lcccccc}
\hline
                & {\bf M1}\textsuperscript{\emph{b}}     & {\bf M2}\textsuperscript{\emph{b}}    & {\bf M3}\textsuperscript{\emph{b}}    & {\bf M4}    & {\bf M5}    & {\bf M6}    \\
\hline
$g_x$           & 1.505 &      1.505 &      1.504 &      1.505 &      1.510 &      1.504 \\
$g_y$           & 1.504 &      1.503 &      1.504 &      1.502 &      1.498 &      1.504 \\
$g_z$           & 1.476 &      1.477 &      1.477 &      1.480 &      1.477 &      1.478 \\
\hline
\end{tabular}
\\
\raggedright
\textsuperscript{\emph{a}} The coordinate system is along the principal axes of the $g$-tensor with the $z$ axis being roughly perpendicular to the ligand plane;
\textsuperscript{\emph{b}} The $g$-tensor was calculated assuming a cationic charge state (see the text).
\label{gtensor}
\end{table}

We evaluate the elements of the $g$-tensor and $B_k^q$ values using the thirteen \emph{ab initio} energy levels and the corresponding eigenfunctions.\cite{chibotaru2012ab,Atanasov2011} The results are shown in Tables \ref{gtensor} and \ref{CF1} (higher order CF parameters are shown in Table S10 in Supporting Information). We use the coordinate system along the principal axes of the $g$-tensor. In this coordinate system, the $z$ axis is roughly perpendicular to the ligand planes. Note that the principal values of the $g$-tensor are close to the ideal Lande $g$-factor value of $3/2$. The calculated $|B_2^0|$ value is the largest for {\bf M5} and the smallest  for {\bf M4}. Due to the absence of any symmetry, however, {\bf M4} and {\bf M5} molecules have significant all-order off-diagonal CF parameters (Table~\ref{CF1}).  In particular, for {\bf M5}, the $B_2^2$ value is about 63\% of the $|B_2^0|$ value, while the $|B_4^4|$ value is similar to the $|B_4^0|$  value. On the other hand, for {\bf M6}, only off-diagonal terms with $q={\pm 4}$ are significant, which is dictated by the molecular $C_4$ symmetry. Including the full set of CF parameters up to sixth order ($k=6$) and the diagonal eighth order term ($B_8^0$), diagonalization of Eq.~(\ref{H1}) reproduces the \emph{ab initio} energy levels up to 0.5~cm$^{-1}$. This indicates that Eq.~(\ref{H1}) is a proper Hamiltonian for the low-energy spectrum of charged TbPc$2$-type SMMs.

\begin{table}
\centering
\caption{\textbf{CF Parameters\textsuperscript{\emph{a},\emph{b}} in cm$^{-1}$ for Tb Ion for {\bf M4}, {\bf M5} and {\bf M6} SMMs} . }
\begin{tabular}{cccc}
\hline
 $B_{k}^q$ & \hspace{0.5cm}{\bf M4}\hspace{0.5cm} & \hspace{0.5cm}{\bf M5}\hspace{0.5cm} & \hspace{0.5cm}{\bf M6}\hspace{0.5cm} \\
\hline
$B_{2}^{-2}$       & -0.12802616 & -0.02332695 & 0.00000240 \\
$B_{2}^{-1}$       & -0.03939183 & -0.00727754 & 0.00001437 \\
$B_{2}^{ 0}$       & -5.05068413 & -5.52638125 & -5.33499223 \\
$B_{2}^{ 1}$       & -0.02663218 & -0.00634158 & -0.00001624 \\
$B_{2}^{ 2}$       &  0.80463902 & 3.45788034 & 0.00003865 \\
\hline
$B_{4}^{-4}$       & -0.00332496 & -0.00213682 & 0.00610215 \\
$B_{4}^{-3}$       &  0.00793656 & -0.00424813 & -0.00000026 \\
$B_{4}^{-2}$       &  0.00410350 & 0.00096348 & -0.00000007 \\
$B_{4}^{-1}$       & -0.00375441 & 0.00049609 & 0.00000289 \\
$B_{4}^{ 0}$       & -0.01406960 & -0.01300444 & -0.01412822 \\
$B_{4}^{ 1}$       & -0.00514061 & 0.00707655 & -0.00000337 \\
$B_{4}^{ 2}$       & -0.00151449 & -0.02369506 & -0.00000004 \\
$B_{4}^{ 3}$       &  -0.00030393 & 0.00188537 & 0.00000259 \\
$B_{4}^{ 4}$       & -0.00079989 & 0.01117566 & 0.00540901 \\
\hline
\end{tabular}
\\
\raggedright
\textsuperscript{\emph{a}} The coordinate system is along the magnetic axes that diagonalize the $g$-tensor ($z$ axis roughly perpendicular to the ligand planes);
\textsuperscript{\emph{b}} Higher order $B_{k}^q$ are provided  in Table S10 in Supporting Information.
\label{CF1}
\end{table}

The lower part of the spectrum (six lowest levels for {\bf M4} and {\bf M6} and four lowest levels for {\bf M5}) is composed of approximate 
doublets, as shown in Fig.~\ref{Spectrum}. The former three doublets correspond to states $M_J=\pm6$, $\pm$5, and $\pm$4, while the latter
two doublets are states $M_J=\pm6$ and $\pm$5, respectively. 
For further level characteristics, see Table S14-S16 in Supporting Information. The transverse CF, however, mixes states with different $|M_J|$ values and breaks 
the degeneracies of the doublets leading to tunnel splitting. It is important to understand tunneling splitting of the levels within the ground multiplet since magnetization can be relaxed via phonon-assisted tunneling. For low-energy levels with large $|M_J|$, the transverse CF has 
a small effect since large powers of its matrix elements are needed to connect states with opposite $M_J$. For high-energy levels with small 
$|M_J|$, the transverse CF becomes more important and tunnel splitting is more pronounced. In fact, for higher part of the spectrum, $|M_J|$ 
ceases to be a good quantum number and the doublet structure disappears. The tunnel splitting and $|M_J|$ mixing for {\bf M5}
are significantly larger than those for {\bf M4} and {\bf M6}. This is because the substantial distortions and curving of the ligand planes 
for {\bf M5} lead to significant transverse CF (Table~\ref{CF1}). The geometrical distortions are shown in the larger spread of $D_{\text{TbN}_{\text{nn}}}$ and $D_{\text{TbC}_{\text{far}}}$ values and of $\theta_{\text{N}_\text{nn}\text{TbN}_\text{nn}}$ and $\theta_{\text{C}_\text{far}\text{TbC}_\text{far}}$ values and in the larger range of
the $Z_{\text{C}^\text{up}}$ and $Z_{\text{C}^\text{down}}$ values in Table 1, compared to the other molecules of interest. For {\bf M5}, the tunnel splitting values are of the order of $10^{-2}$, $10^{-1}$, and 10~cm$^{-1}$ for the ground state and the first- and second-excited states, respectively (Table~\ref{Scales}). For {\bf M6}, the $C_4$ symmetry allows mixing of $M_J$ levels with $\Delta M_J$$=$$\pm 4$ only. This explains the significant tunnel splitting only for a pair of states 5 and 6 and a pair of states 9 and 13 (Table~\ref{Scales} and Table S2 in Supporting Information).

The calculated ZFS and MAB values are shown in Table \ref{Scales}. We find that ZFS lies in the range of 289-292~cm$^{-1}$ for the anionic molecules ({\bf M4} and {\bf M5}), while it is somewhat larger for the cationic molecule ({\bf M6}). As seen from Fig. \ref{Spectrum}, ZFS for {\bf M6} is similar to ZFS for other cationic TbPc$_2$ or TbPcNc molecules. Therefore, we conclude that ZFS does not depend on ligand type and geometry details, and it only shows a weak dependence on oxidation number. This reflects the fact that ZFS characterizes the lower part of the energy spectrum where the transverse CF has a small effect. However, higher-energy part of the spectrum where the transverse CF plays a significant role reveals a much stronger dependence on the ligand type and geometry details, as shown in Fig.~\ref{Spectrum}.

Our calculated results for {\bf M5} molecule can be directly compared with similar calculations from Ref.~\citenum{Ungur2017} where the same experimental geometry was used. ZFS and MAB reported in Ref.~\citenum{Ungur2017} are 308~cm$^{-1}$ and 809~cm$^{-1}$, respectively. While ZFS is reasonably close to our value (292~cm$^{-1}$), the MAB differs more significantly from our value (770~cm$^{-1}$). We check that both the choice
of the active space and the effect of higher spin-free states are not responsible for the significant difference (see Supporting Information). The most likely reason is the basis set difference between our study and Ref.~\citenum{Ungur2017}.

On the experimental side, the Tb CF parameters for [TbPc$_2$]$^{-}$TBA$^+$ \cite{Ishikawa2003} estimated by Ishikawa {\it et al.}\cite{Ishikawa2003InorgChem} are widely used in the community. These CF parameters were obtained by fitting both the experimental nuclear magnetic resonance (NMR) shift and magnetic susceptibility data to a ligand-field model, assuming perfect C$_4$ symmetry. The estimated CF parameters are listed in Thiele {\it et al.}\cite{Thiele2014}. In this estimate, three uniaxial terms $B_2^0$, $B_4^0$, and $B_6^0$ as well as only one transverse term $B_4^4$ were considered. The estimated values are $B_2^0$$=$$-$4.18193, $B_4^0$$=$$-$0.02790, $B_4^4=$0.00122, and $B_6^0$$=$$-$0.00004~cm$^{-1}$. Since X-ray crystallography data was not reported and perfect C$_4$ symmetry was
assumed in Ishikawa {\it et al.},\cite{Ishikawa2003InorgChem} our calculated CF parameters cannot be directly compared to their fitted 
values. Nonetheless, it is worthwhile providing some remarks. Note that {\bf M4} molecule has the same counter cation without bulky 
dilution molecules as in Ishikawa {\it et al.}\cite{Ishikawa2003InorgChem}. Thus, we discuss major differences between our calculated CF 
parameters for {\bf M4} molecule with those by Ishikawa {\it et al.}\cite{Ishikawa2003InorgChem}. For {\bf M4} molecule, the $|B_2^0|$ value 
is somewhat greater than the fitted value, while the $|B_4^0|$, $|B_4^4|$, and $|B_6^0|$ values are 
comparable to the fitted values. See Table~\ref{CF2} and Table S10 in Supporting Information. The major difference is that we find 
large low-order transverse CF parameters such as $|B_2^{q\neq0,2}|$, $|B_4^{q\neq0,4}|$ and $|B_6^{q\neq0}|$, whereas the literature 
considered only one transverse CF parameter $B_4^4$. As a result, our result predicts a much larger tunnel splitting and qualitatively 
different level characteristics above the first-excited states.

\begin{table}
\centering
\caption{\textbf{Measured effective magnetic anisotropy barrier (in cm$^{-1}$) for TbPc$_2$-like single-molecule magnets}}
\begin{tabular}{lccccccc}
\hline
Molecule & TbPc$_2$\textsuperscript{\emph{a}} & TbPc$_2$\textsuperscript{\emph{b}}   & TbPc$_2$\textsuperscript{\emph{b}}   & 
TbL$_2$\textsuperscript{\emph{c,d}} &  TbL$_2$\textsuperscript{\emph{c,d}} &
TbPcNc\textsuperscript{\emph{e}}    & TbPcNc\textsuperscript{\emph{e}} \\
\hline
Charge   & -1       & -1         & -1         & -1        & +1       & +1        & 0      \\
Dilution & Yes      & No         & Yes        & No        & No       & No        & No     \\
MAB  & 260 & 584      & 641        & 594       & 550      & 584       & 342    \\
Relevance & n/a     & {\bf M5}   & {\bf M5}   & n/a       & n/a      & {\bf M6}  & {\bf M3} \\
\hline
\end{tabular}
\\
\raggedright
\textsuperscript{\emph{a}} Ref.~\citenum{Ishikawa2003};
\textsuperscript{\emph{b}} Ref.~\citenum{Branzoli2009};
\textsuperscript{\emph{c}} L=Pc(OEt)$_8$;
\textsuperscript{\emph{d}} Ref.~\citenum{Takamatsu2007};
\textsuperscript{\emph{e}} Ref.~\citenum{Katoh2018};
\label{Ueff}
\end{table}

Table~\ref{Ueff} shows the experimental effective barrier for several
TbPc$_2$-like molecules. The measured barrier for {\bf M6} was reported, considering Raman and Orbach processes, while the barrier for a diluted crystal of neutral TbPcNc ({\bf M3}) molecules was obtained, considering Orbach and quantum tunneling processes.\cite{Katoh2018} It is interesting 
to compare the measured barrier for {\bf M5} with and without bulky diamagnetic dilution molecules.\cite{Branzoli2009} These experimental values are much larger than the earlier reported values from Refs.~\citenum{Ishikawa2003} and \citenum{Ishikawa2004}, for anionic TbPc$_2$ molecular crystals. Comparison with the experimental data indicates that for {\bf M6} molecule, the experimental barrier is close
to states 11 and 12, and for {\bf M5} molecule, the range of the experimental barrier falls on states 10 and 11. See Table S2 in Supporting
Information.

\subsection{Neutral molecules}

\begin{figure}
\centering
\includegraphics[width=1.0\linewidth]{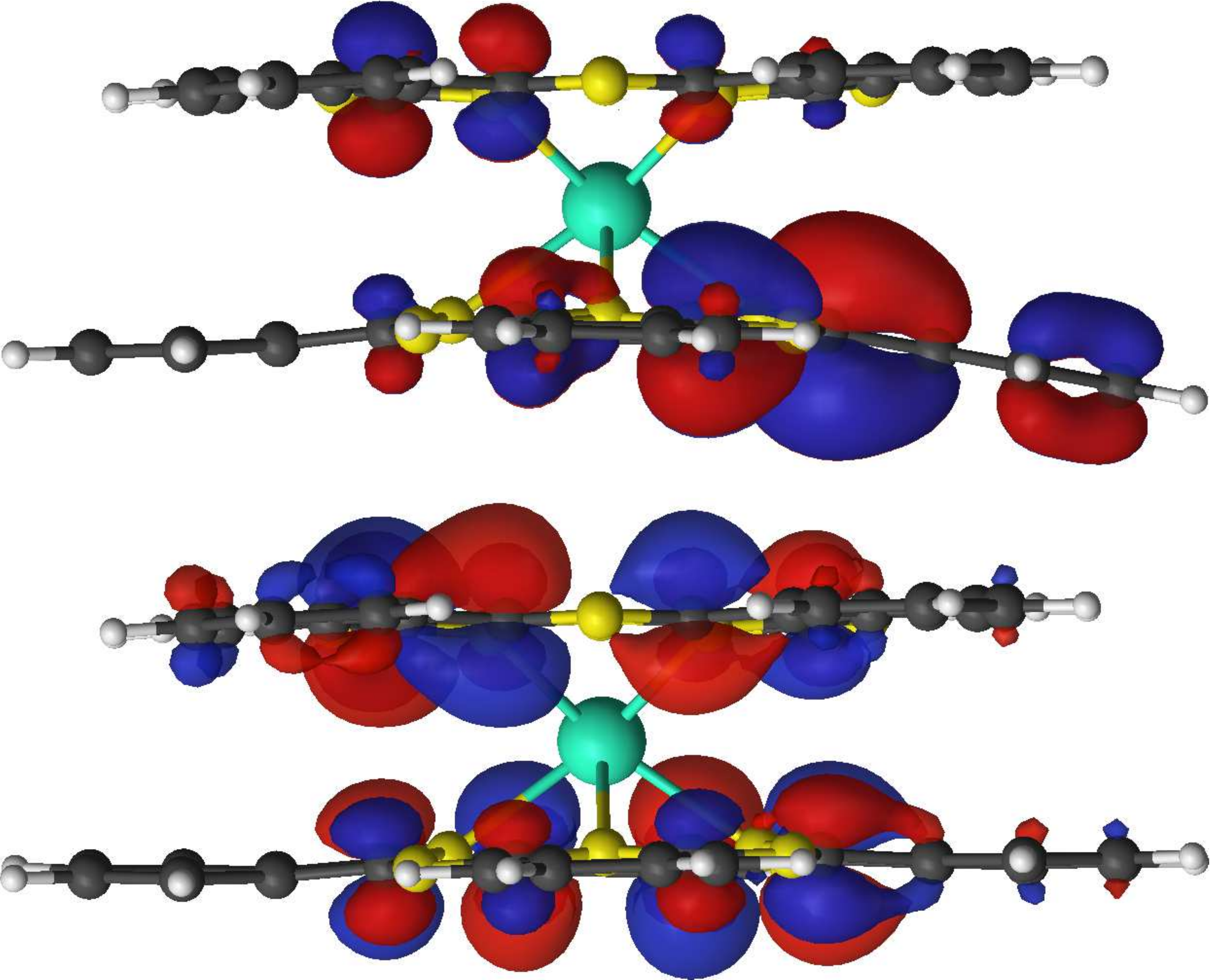}
\caption{Graphical representation of SOMO (L2) obtained from scalar relativistic CASSCF calculations for neutral {\bf M1} molecule (top) and for a flattened {\bf M1} molecule (bottom). Red and blue regions represent opposite phases of the orbital wave function. The figure is prepared using the Luscus software.\cite{Luscus}}
\label{LigandOrbital}
\end{figure}

\begin{figure}
\centering
\includegraphics[width=1.0\linewidth]{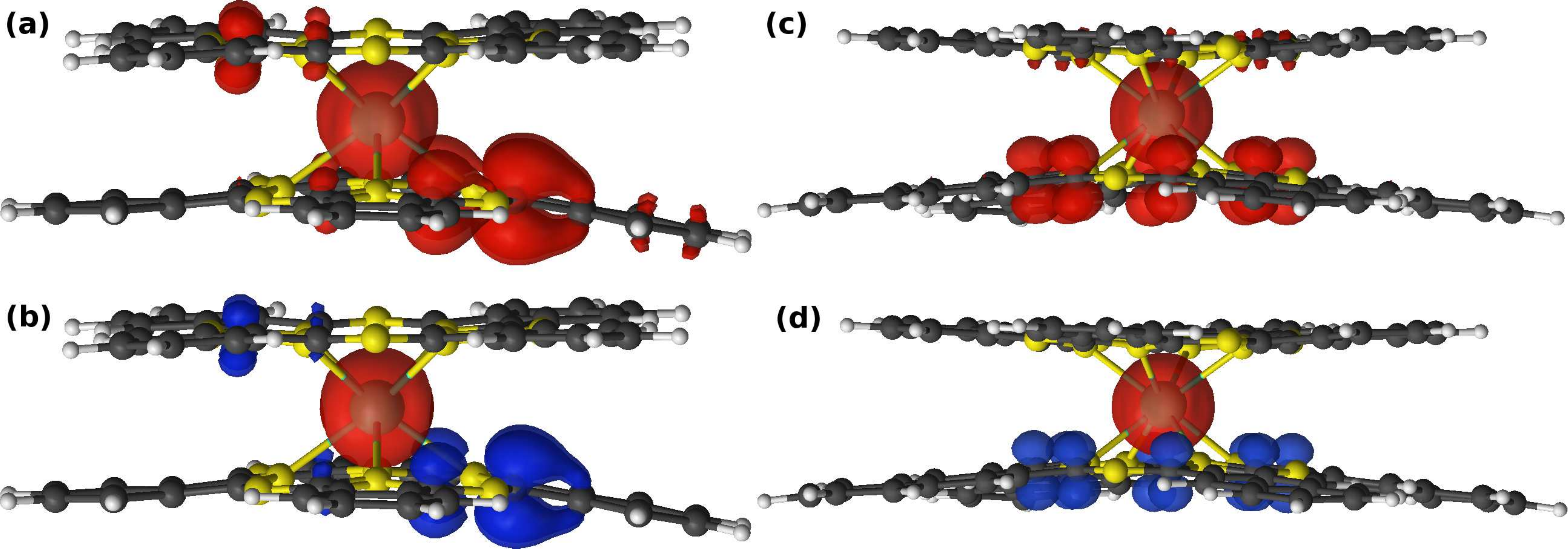}
\caption{Spin density from scalar relativistic CASSCF calculations for {\bf M1} molecule for $S_{\text{tot}}=7/2$ (a) and $S_{\text{tot}}=5/2$ (b) and for {\bf M3} molecule for $S_{\text{tot}}=7/2$ (c) and $S_{\text{tot}}=5/2$ (c). Red and blue colors denote spin-up and spin-down density, respectively. The figure is prepared using the Luscus software.\cite{Luscus}}
\label{SpinDensity}
\end{figure}

Neutral TbPc$_2$-type molecules ({\bf M1}, {\bf M2} and {\bf M3}) have much richer electronic structure than their charged counterparts due to presence of an extra unpaired electron that is expected to reside in the ligands. As in the case of the charged molecules, seven of the eleven active orbitals are Tb 4$f$-like (Fig. S1 in Supporting Information). These orbitals are similar for all considered molecules and occupation of each of these orbitals is roughly 1.14 consistent with eight electrons occupying the 4$f$ shell. The remaining four active orbitals (L1-L4) are ligand orbitals and are shown in Figs. S2-S4 in Supporting Information. For all neutral molecules L1 is almost doubly occupied, L2 is singly occupied (SOMO) while L3 and L4 have small occupation numbers. For {\bf M3} molecule, the SOMO (L2) has top-down plane asymmetry due to different ligand type but shows in-plane symmetry. Interestingly, for {\bf M1} and {\bf M2} molecules, despite the same ligand type, the SOMO is highly asymmetric with large weight on the bottom Pc plane only (Fig. S2 and S3). This is due to the asymmetric curvature within the two Pc planes, as listed in Table 1. Compare the $Z_{\text{C}^\text{up}}$ and $Z_{\text{C}^\text{down}}$ values in Table 1. In order to check on this, we enforce mirror symmetry about the $xy$ plane ($\sigma_h$) in {\bf M1} molecule and compute the ligand orbitals 
in the active space. For this flattened {\bf M1} molecule, we find that top-bottom plane symmetry is more or less restored in the SOMO. Compare Fig.~\ref{LigandOrbital}(a) with (b).

The spin of the ligand electron can be either parallel ($S_{\text{tot}}=7/2$) or antiparallel ($S_{\text{tot}}=5/2$) to the Tb spin. The calculated spin density for both cases is shown in Fig. \ref{SpinDensity} for {\bf M1} and {\bf M3} molecules. The Tb spin density is localized in the vicinity of Tb ionic core. For {\bf M1}, the ligand spin density is mostly shared by inner carbon atoms in one side of the bottom Pc plane.  This is consistent with the SOMO (Fig. \ref{LigandOrbital}(a)) that is primarily delocalized on these carbon atoms. The same behavior is observed for {\bf M2} molecule (not shown). For {\bf M3}, consistently with SOMO, the majority of the ligand spin density is symmetrically shared by the inner carbon atoms from the Nc ligand.
Table S1 in Supporting Information shows the calculated seven lowest spin-free states for both values of $S_{\text{tot}}$ for different neutral molecules. For all systems, the parallel configuration of the Tb and ligand spin ($S_{\text{tot}}=7/2$) has a lower energy than the antiparallel configuration ($S_{\text{tot}}=5/2$). 

\begin{figure}
\centering
\includegraphics[width=0.6\linewidth]{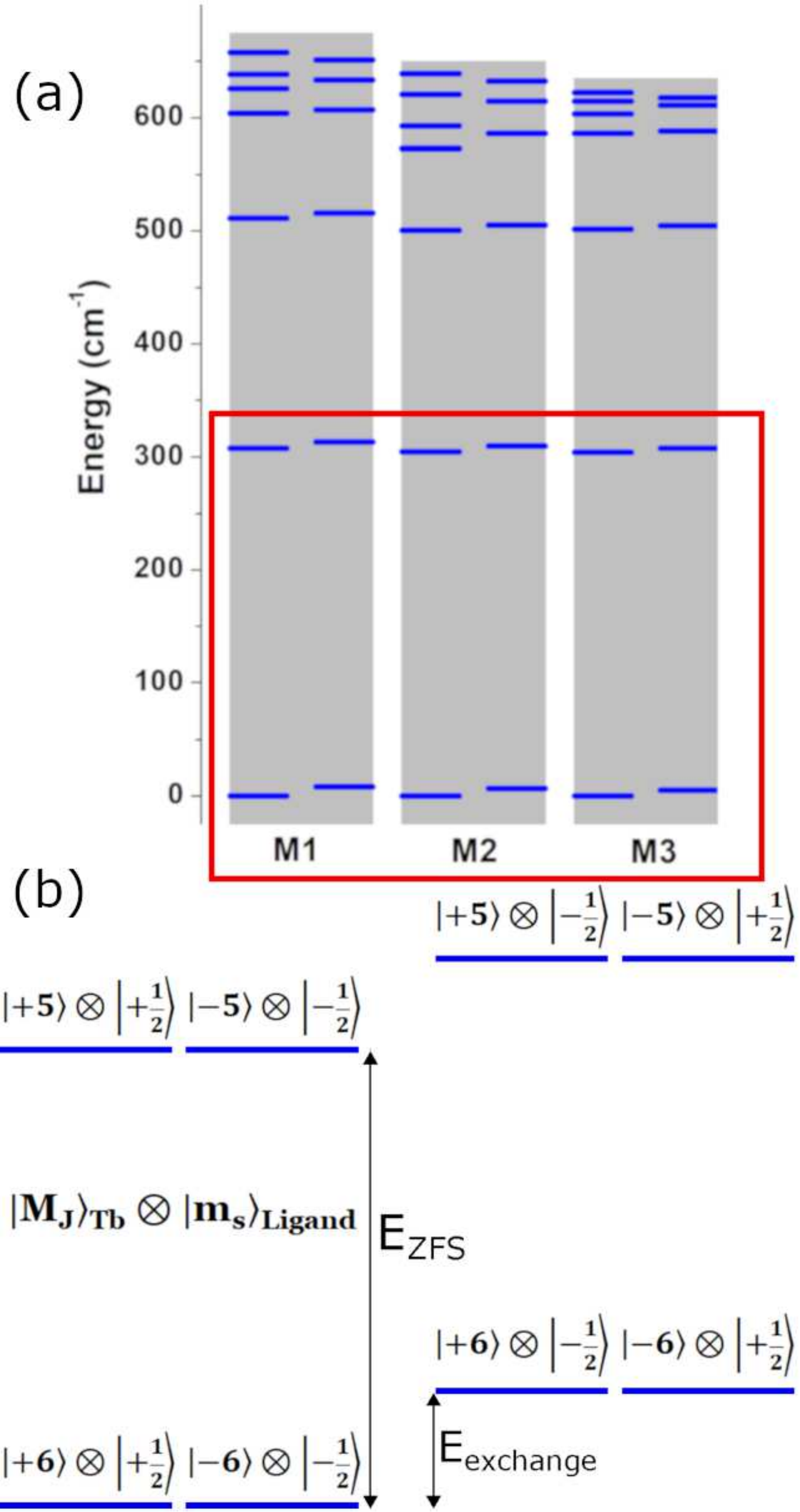}
\caption{(a) Calculated low-energy spectrum for {\bf M1}, {\bf M2}, and {\bf M3} neutral molecules. Each line represents a Kramers doublet. The red rectangle denotes the low-energy part of the spectrum that is schematically illustrated in panel (b).}
\label{Spectrum2}
\end{figure}

\begin{table}
\centering
\caption{\textbf{CF Parameters\textsuperscript{\emph{a},\emph{b}} in cm$^{-1}$ for Tb Ion for {\bf M1}, {\bf M2} and {\bf M3} SMMs}}
\begin{tabular}{cccc}
\hline
 $B_{k}^q$ & \hspace{0.5cm}{\bf M1}\textsuperscript{\emph{c}}\hspace{0.5cm} & \hspace{0.5cm}{\bf M2}\textsuperscript{\emph{c}}\hspace{0.5cm} & \hspace{0.5cm}{\bf M3}\textsuperscript{\emph{c}}\hspace{0.5cm} \\
\hline
$B_{2}^{-2}$       &  0.00000001 & -0.04824674 & -0.00000478 \\
$B_{2}^{-1}$       & -0.00000003 &  0.03000187 &  0.00000119 \\
$B_{2}^{ 0}$       & -5.93432163 & -5.68334414 & -5.65157592 \\
$B_{2}^{ 1}$       & -0.00444594 & -0.04186354 & -0.00001973 \\
$B_{2}^{ 2}$       &  0.32444131 &  0.44453996 &  0.00019387 \\
\hline
$B_{4}^{-4}$       &  0.00000000 & -0.01188166 & -0.00302990 \\
$B_{4}^{-3}$       &  0.00000000 & -0.00089983 &  0.00000000 \\
$B_{4}^{-2}$       &  0.00000000 &  0.00162901 &  0.00000014 \\
$B_{4}^{-1}$       &  0.00000000 & -0.00034396 &  0.00000135 \\
$B_{4}^{ 0}$       & -0.01336535 & -0.01369266 & -0.01380654 \\
$B_{4}^{ 1}$       & -0.00043523 & -0.00806695 & -0.00000275 \\
$B_{4}^{ 2}$       & -0.00143680 & -0.00160371 & -0.00000081 \\
$B_{4}^{ 3}$       &  0.00160576 & -0.00288249 & -0.00000487 \\
$B_{4}^{ 4}$       & -0.00027466 & -0.01198852 & -0.00720075 \\
\hline
\end{tabular}
\\
\raggedright
\textsuperscript{\emph{a}} The coordinate system is along the magnetic axes that diagonalize the $g$-tensor ($z$ axis roughly perpendicular to the ligand planes);
\textsuperscript{\emph{b}} Higher order $B_{k}^q$ are provided  in Table S10 in Supporting Information;
\textsuperscript{\emph{c}} The CF parameters were calculated assuming a cationic charge state (see the text).
\label{CF2}
\end{table}

Since neutral TbPc$_2$ and TbPcNc molecules have an odd number of electrons, the Kramers theorem dictates that all electronic levels are at least doubly degenerate. The low symmetry of the considered molecules prevents from appearance of higher degeneracy. The low-energy part of the calculated spectra for {\bf M1}, {\bf M2}, and {\bf M3} molecules consists of a group of thirteen Kramers doublets (Fig. \ref{Spectrum2} and Table~S3 in Supporting Information).

The thirteen lowest Kramers doublets result from Tb $J=6$ multiplet that is coupled with the unpaired ligand electron spin. The next electronic 
level belongs to the first-excited multiplet and it lies more than 1425~cm$^{-1}$ higher than the highest level in the first multiplet 
(Table \ref{Scales}). Therefore, for analysis of the magnetic properties we can focus only on the thirteen lowest doublets. Such an energy spectrum 
can be described by the following generalization of Eq.~(\ref{H1}) in order to include the ligand spin:

\begin{equation}
\hat{H}=\sum_{kq}B_k^q\hat{O}_k^q(\hat{\mathbf{J}}) +J_{\text{ex}}\hat{\mathbf{s}}\cdot\hat{\mathbf{J}}-\mu_B\mathbf{B}\cdot\left(\mathbf{g}\cdot\hat{\mathbf{J}}+g_e\hat{\mathbf{s}}\right),
\label{H2}
\end{equation}
where we introduce Heisenberg exchange coupling $J_{\text{ex}}$ between the ligand spin ($\hat{\mathbf{s}}$) and the Tb total angular momentum operators. In addition, the Zeeman term is generalized to describe interaction of the ligand spin with magnetic field ($g_e\approx2$ is a free-electron $g$-factor). Note that in the presence of strong SOI, anisotropic and antisymmetric exchange couplings may play an important role and rigorous treatment of exchange interaction requires more general formalism.\cite{Malrieu2014,Iwahara2015,Vieru2016} We expect, however, that the simple isotropic Heisenberg form in Eq.~(\ref{H2}) is a good approximation for our molecules since the exchange interaction is very small. The anisotropic and antisymetric exchanges (as well as higher order spin interactions) are, in general, smaller than the isotropic exchange and, therefore should not play a significant role for TbPc$_2$-type molecules. As discussed below, we find that Hamiltonian~(\ref{H2}) provides a good representation of the low-energy spectrum of the neutral molecules. 

First-principles evaluations of the Tb $B_k^q$ coefficients and the $g$-tensor elements are not straightforward for the neutral molecules. As discussed in Sec.~3, we obtain the Tb $B_k^q$ coefficients and the $g$-tensor by using the calculated low-energy spectra (Fig.~\ref{Spectrum}(a)-(c) and Table S2 in Supporting Information) of the neutral molecular geometries with one electron removed. Here we assume that the contribution of the unpaired ligand electron to the Tb CF and $g$-tensor is negligible. Later we show that this is indeed a valid assumption. The calculated elements of the $g$-tensor and the Tb CF parameters are shown in Tables~ \ref{gtensor} and \ref{CF2} (see Table S10 in Supporting Information for a full set of $B_k^q$). Note that as in the case of {\bf M4}, {\bf M5} and {\bf M6} molecules, we use a coordinate system along the principal axes of the $g$-tensor for which the $z$ axis (easy axis) points approximately in the direction perpendicular to the ligand planes. We find all the principal values of the $g$-tensor being close to the ideal Lande $g$-factor value of $3/2$ (Table~\ref{gtensor}). The calculated $|B_2^0|$ value for {\bf M1} is a bit larger than that for {\bf M2} and {\bf M3}, which is reflected in the largest ZFS among the three neutral molecules. The calculated $|B_2^0|$ value for {\bf M1}, {\bf M2}, and {\bf M3} is consistently larger than that for the charged molecules. Compared to the charged TbPc$_2$ molecules, the degree of the structural distortion is much less for the neutral TbPc$_2$ molecules. See $D_{\text{TbC}_{\text{far}}}$, $Z_{\text{C}^\text{up}}$, and $Z_{\text{C}^\text{down}}$ values in Table 1. This explains overall smaller transverse CF parameters in the neutral TbPc$_2$ molecules than in the charged TbPc$_2$ molecules.

In order to evaluate the exchange coupling $J_{\text{ex}}$, we fit the {\it ab-initio} energies of the lowest thirteen doublets to eigenvalues of Eq.~(\ref{H2}) with a fitting parameter $J_{\text{ex}}$. For all molecules, we obtain a high-quality fit.
This indicates that the method we use for evaluation of CF parameters is reliable and that Eq.~(\ref{H2}) provides a reasonable description of the low-energy spectrum of neutral TbPc$_2$-type molecules. We find a small ferromagnetic exchange coupling with $J_{\text{ex}}\approx 0.6-0.8$~cm$^{-1}$ (Table \ref{Scales}). An increase of the active space or the atomic basis sets does not affect this number significantly (see Tables S4 and S7 in Supporting Information). The insensitivity of $J_{\text{ex}}$ to ligand type and synthesis process indicates that the exchange coupling does not depend on details of molecular geometry much as long as structural changes are moderate. This conclusion is only applied to molecules in crystals or films on substrates rather than within single-molecule transistors where the molecules experience much stronger structural changes.

Using the pseudospin Hamiltonian, Eq.~(\ref{H2}), we can determine the characteristics of energy levels shown in Fig. \ref{Spectrum2}. For all considered molecules, four lowest Kramers doublets have a well-defined $|M_J|$ value. For each pair, the lower (higher) energy level has the ligand spin parallel (antiparallel) to the Tb angular momentum. Within each Kramers doublet, it is convenient to choose such a basis set in which each Kramers partner state is characterized by $M_J$ and $m_s$ quantum numbers. Since Kramers partner states are related by time reversal symmetry, they have opposite values of $M_J$ and $m_s$. If we focus on the Kramers partner state with positive $M_J$, we find ($M_J=6$, $m_s=1/2$) for the
ground state level (level 1), ($M_J=6$, $m_s=-1/2$) for the first-excited level (level 2), ($M_J=5$, $m_s=1/2$) for the second-excited level 
(level 3), ($M_J=5$, $m_s=-1/2$) for the third-excited level (level 4) (Fig.~\ref{Spectrum2}). For the most symmetric {\bf M3} molecule, 
level 5 consists of a majority of ($M_J=4$, $m_s=1/2$) and small contributions from ($M_J=-4$, $m_s=1/2$) and ($M_J=0$, $m_s=1/2$), while 
level 6 consists of a majority of ($M_J=4$, $m_s=-1/2$) with small contributions from ($M_J=-4$, $m_s=-1/2$), ($M_J=-4$, $m_s=1/2$), ($M_J=0$, $m_s=-1/2$) and ($M_J=3$, $m_s=1/2$).

For the neutral molecules, we define ZFS as an energy difference between $|M_J|=6$ and $|M_J|=5$ levels with the same direction of ligand spin with respect to the Tb total angular momentum (values for parallel and antiparallel configurations are very close). As seen in Table \ref{Scales} and Fig.~\ref{Spectrum2}, ZFS is similar for {\bf M1}, {\bf M2} and {\bf M3} molecules. This indicates that ZFS is not sensitive to the ligand type and geometry details, similarly to $J_{\text{ex}}$. However, the ZFS for the neutral molecules is somewhat larger than the values for the anionic and cationic molecules. Higher-energy part of the spectrum shows some dependence on ligand type and geometry details but we have less sensitivity than in the case of charged molecules.

We now compare our calculated CASSCF-RASSI-SO results to other theoretical calculations and experimental data. Regarding $J_{\text{ex}}$, a similar value to our value has been obtained from CASSCF-RASSI-SO calculations for a neutral TbPc$_2$ molecule adsorbed on a Ni substrate using a DFT-optimized atomic structure.\cite{marocchi2016relay} Compared to the experimental $J_{\text{ex}}$ from electron paramagnetic resonance spectra for a crystal of {\bf M1} molecules,\cite{Komijani2018} the sign of our calculated $J_{\text{ex}}$ agrees with experiment but the magnitude is a bit larger than the experimental value. Experiments on TbPc$_2$-based single-molecule transistors have shown both antiferromagnetic~\cite{Thiele2014} and ferromagnetic \cite{urdampilleta2015magnetic} couplings between the ligand spin and the Tb spin. These seemingly conflicting experimental results are not 
surprising, considering the small energy scale of $J_{\text{ex}}$ and possible large configurational changes of the ligand planes of the TbPc$_2$ molecule bridged between gold electrodes in single-molecule transistor set-ups. Experimental data for the effective energy barrier is rare for 
neutral TbPc$_2$ molecules. There exists an experimental report on a neutral TbPcNc molecular crystal diluted with YPcNc.\cite{Katoh2018} In this 
case, the experimental barrier is 342~cm$^{-1}$ (Table 5). The measured value seems to be close to our calculated ZFS for {\bf M3} molecule. 
See Tables 2 and 6.

\section{\label{sec:conc}Conclusions}

We investigate electronic structure and magnetic properties of six different TbPc$_2$ and TbPcNc molecules in different charge states using relativistic multireference methods and pseudospin Hamiltonian technique. For the charged molecules, we evaluate CF parameters and $g$-tensor elements by projecting the low-energy spectrum onto effective $J=6$ pseudospin. For the neutral molecules, we consider exchange coupling of the Tb magnetic moment with the ligand spin and extract Tb CF parameters and $g$-tensor elements by separating the Tb ground multiplet from the unpaired ligand electron spin-flip states using artificial oxidation. The key findings are as follows:

\begin{itemize}

\item For the neutral molecules, the exchange coupling constant between the Tb magnetic moment and the ligand spin does not depend much on ligand type and geometry details. 
This result is valid as long as molecular structures are more or less controlled such as in crystals or layers on substrates. 

\item Geometry details and ligand type do not affect ZFS.

\item ZFS weakly depends on oxidation number. The neutral
molecules have somewhat higher ZFS than the charged molecules.

\item The higher-energy levels and associated tunnel splittings strongly depend on ligand type, oxidation number, and geometry details.

\item Comparison to experimental effective barrier suggests that in some cases higher-energy levels rather than just ZFS may contribute to the magnetization relaxation through phonon-assisted tunneling.

\end{itemize}

These results provide insights in separating the effects intrinsic to individual molecules from extrinsic effects on magnetization relaxation and in interpretation of reported experimental data and stimulating new experiments on TbPc$_2$-type molecules.

\begin{acknowledgement}
This work was funded by the Department of Energy (DOE) Basic Energy Sciences (BES) grant No DE-SC0018326. Computational support by Virginia Tech ARC and San Diego Supercomputer Center (SDSC) under DMR060009N. We also thank Dr. Benjamin Pritchard for helpful discussion and insight.
\end{acknowledgement}

\begin{suppinfo}
The Supporting Information is available free of charge: Molecular orbitals, spin-free energies, numeric data for energy levels, basis set and active space dependence, definitions of extended Stevens operators, full set of crystal field parameters, and character of electronic levels.
\end{suppinfo}

\bibliography{refs}




\end{document}